\documentclass[12pt]{iopart}
\pdfoutput=1
\eqnobysec
\usepackage{iopams}
\usepackage{braket}
\usepackage{cite}
\usepackage{iopams}
\usepackage{setstack}
\usepackage[usenames,dvipsnames]{color}
\definecolor{myorange}{RGB}{199.24, 87.48, 47.80}
\usepackage[colorlinks,bookmarks=false,citecolor=NavyBlue,linkcolor=OliveGreen,urlcolor=blue]{hyperref}
\usepackage{graphicx}
\newcommand{\be}{\begin{equation}}
\newcommand{\ee}{\end{equation}}
\newcommand{\bea}{\begin{eqnarray}}
\newcommand{\eea}{\end{eqnarray}}
\newcommand{\ba}{\begin{aligned}}
\newcommand{\ea}{\end{aligned}}

\usepackage{relsize}
\usepackage{tikz}
\usetikzlibrary{shapes.arrows, fadings}
\tikzfading[name=fade right,
  left color=transparent!0, right color=transparent!100]
  \tikzfading[name=fade left,
  left color=transparent!100, right color=transparent!0]
\definecolor{myorange}{RGB}{199.24, 87.48, 47.80}
\usetikzlibrary{decorations.markings}

\begin{document}
\title{Entanglement evolution and generalised hydrodynamics: noninteracting systems }
\author{Bruno Bertini}
\address{Department of physics, FMF, University of Ljubljana, Jadranska 19, SI-1000 Ljubljana, Slovenia}
\author{Maurizio Fagotti}
\address{LPTMS, CNRS, Univ. Paris-Sud, Universit\'e Paris-Saclay, 91405 Orsay, France}
\author{Lorenzo Piroli}
\address{SISSA and INFN, via Bonomea 265, 34136 Trieste, Italy}
\author{Pasquale Calabrese}
\address{SISSA and INFN, via Bonomea 265, 34136 Trieste, Italy}
\address{International Centre for Theoretical Physics (ICTP), I-34151, Trieste, Italy}

\begin{abstract}
The large-scale properties of homogeneous states after quantum quenches in integrable systems have been successfully described 
by a semiclassical picture of moving quasiparticles. 
Here we consider the generalisation for the entanglement evolution after an inhomogeneous quench in noninteracting systems 
in the framework of generalised hydrodynamics. 
We focus on the protocol where two semi-infinite halves are initially prepared in different states and then joined together,
showing that a proper generalisation of the quasiparticle picture leads to exact quantitative predictions. 
If the system is initially prepared in a quasistationary state, we find that the entanglement entropy is additive  
and it can be computed by means of generalised hydrodynamics. 
Conversely, additivity is lost when the initial state is not quasistationary; yet the entanglement entropy in the large-scale limit can be exactly predicted in the quasiparticle picture, provided that the initial state is low entangled. 
\end{abstract}

\maketitle

\section{Introduction}  %

The puzzle about the emergence of statistical mechanics and thermodynamics from a microscopic quantum theory 
has recently found a natural basis in the context of bipartite entanglement. 
Indeed, it has been established that non-equilibrium isolated pure systems may relax locally to thermodynamic 
ensembles (see e.g. the reviews \cite{ge-15,cem-16,ef-16,ViRi16}).
Thus, given a subsystem $A$ and a non-equilibrium pure state $|\psi\rangle$, with reduced density matrix 
$\boldsymbol \rho_A\equiv{\rm Tr}_{\bar A}\,|\psi\rangle\langle\psi|$,
the corresponding entanglement entropy
\begin{equation}
S_A\equiv-{\rm Tr}\,\boldsymbol\rho_A\ln\boldsymbol \rho_A\,,
\label{Sdef}
\end{equation}
does not only measure quantum correlations \cite{rev-enta}, but also provides a direct bridge between the microscopic quantum 
world and thermodynamics. 
From this point of view, the entanglement entropy $S_A$ represents the natural 
candidate for a definition of a thermodynamic entropy in the limit of large times after a quantum 
quench \cite{dls-13,CoKC14,bam-15,kaufman-2016,AlCa17,nahum-17} (see 
also \cite{Polk11,SaPR11,DKPR16,PVCR17} for some alternative definitions of entropy).

Practically, qualitative and quantitative predictions for the spreading of bipartite entanglement can be obtained within the semiclassical picture introduced in \cite{CCsemiclassics}. This picture leads to \emph{exact} results for the time evolution of the entanglement entropy in homogeneous settings, which have been tested in many analytical and numerical computations \cite{FaCa08,CCsemiclassics,nahum-17,dmcf-06,lauchli-2008,ep-08,hk-13,Gura13,Fago13,ctc-14,nr-14,buyskikh-2016,cotler-2016,MBPC17,kctc-17,AlCa17,mkz-17,p-18,fnr-17,ckt-18}.
Now, thanks to the recent development of generalised hydrodynamics (GHD)~\cite{CaDY16,BCDF16}, it is finally possible to 
study whether and how the semiclassical picture can be modified to incorporate inhomogeneities. Generalised hydrodynamics is an integrability-based approach to analytically investigate time evolution in the presence of inhomogeneities, even for interacting systems. 
The paradigmatic example of a non-trivial situation described by GHD is the junction of two semi-infinite systems prepared in 
macroscopically different states. At late times $t$,   
the system becomes locally quasistationary and the expectation values of observables at distance $x$ from the junction are functions only of the ratio $\zeta=x/t$~\cite{BF16}, not of $x$ and $t$ separately.

\begin{figure}[t]
\begin{center}
	\includegraphics[scale=1.05]{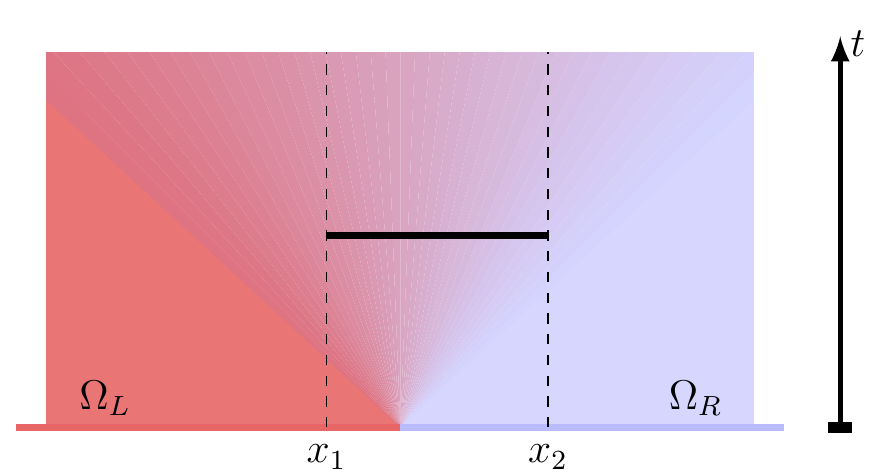}
	\caption{
		Two semi-infinite chains are prepared in different states and then suddenly joined together. We study the entanglement between an interval $A=[x_1,x_2]$ and the rest of the system.
}
	\label{fig:protocol}
\end{center}	
\end{figure}

Following Refs. \cite{CaDY16,BCDF16}, several studies have focused on the behavior of local observables and the spreading of correlations \cite{DoYo17,DDKY17,IlDe17,PDCB17,Bulchandani-17,BVKM17,cdv-17,lzp-17,Fago17-higher,BePC18,BePi17,bdwy-18,mvc-18, F:current}, while little attention has been devoted to the entanglement dynamics, with only few exceptions \cite{ADSV16,SDCV17,Alba17,em-18,WWR18}. In this work we make a step forward in this direction and study the leading behaviour of the entanglement entropy of a single interval $A=[x_1,x_2]$
as a function of the time in the asymptotic limit of large lengths of the subsystems. 
We consider the protocol where two semi-infinite systems, initially prepared in different macrostates $\Omega_L$ and $\Omega_R$, are suddenly joined together at their boundary and left to evolve with a translationally invariant Hamiltonian, see Fig.~\ref{fig:protocol}.   
We focus on two prototypical examples: 
(a) $\Omega_L$ and $\Omega_R$ are two different stationary states of $H$; 
(b) $\Omega_L$ and $\Omega_R$ are two nonstationary low entangled states. 
The first situation can be realised by  joining together two halves prepared in equilibrium at different temperatures. 
The second situation typically arises when one has a Hamiltonian $H(h)$, depending on a global parameter $h$, and
the two halves are initially prepared in the ground state of $H(h)$ but with different values of $h$ (an inhomogeneous global quench~\cite{BCDF16,chl-08}).

We focus on systems of noninteracting fermions, i.e. systems whose Hamiltonian can be (effectively) written as a sum of independent modes  
\be
 H= \sum_{k} \varepsilon(k)  b^\dag_{k}  b^{\phantom{\dag}}_{k} =\sum_{k} \varepsilon(k) \hat n_{k}\,.
\label{eq:Hgen}
\ee 
Here $b^\dag_k,  b^{\phantom{\dag}}_k$ are fermionic creation and annihilation operators,  $\hat n_k\equiv  b^\dag_k b^{\phantom{\dag}}_k$ is the number operator, and 
$\varepsilon(k)$ is the dispersion relation. To simplify the discussion, we assume reflection symmetry ($\varepsilon(k)=\varepsilon(-k)$) and the absence of additional ``hidden symmetries''  (this excludes super-integrable systems like the XY model in zero magnetic field, see Ref.~\cite{F:current}).
In the thermodynamic limit, the sum over $k$ becomes an integral defined on $[-\pi,\pi]$ for lattice models and on the real line for continuous systems.
Our general results will be tested against exact numerical data in the paradigmatic example of the quantum Ising chain
\be
H=- \sum_{j=-L/2+1}^{L/2-1} \boldsymbol \sigma_\ell^x\boldsymbol \sigma_{\ell+1}^x-h\sum_{j=-L/2+1}^{L/2}\boldsymbol \sigma_\ell^z\,,
\label{eq:H}
\ee
where $\{\boldsymbol \sigma_\ell^\alpha\}_{\alpha=x,y,z}$ are Pauli matrices, $h$ is the transverse field, and we adopted open boundary conditions.
This Hamiltonian can be written in the form \eref{eq:Hgen} with ${\varepsilon(k)=2\sqrt{1+h^2-2h\cos k}}$\footnote{Note that, in the thermodynamic limit, the bound state for $|h|<1$ is sent to infinity.}.
Numerical results for the time evolution of the entanglement entropy may be obtained by standard methods
based on the Wick's theorem~\cite{CCsemiclassics, FaCa08, pe-09, lrv-03}; a short summary is given in~\ref{app:num}.

Note that, in general, one needs a non-local transformation (known as Jordan-Wigner transformation) to map the Hamiltonian of the specific spin chain considered to the form \eref{eq:Hgen}. The  non-locality of such transformation does not affect the bipartite entanglement as long as the initial state is  Gaussian and the subsystem is a block of contiguous spins. If the subsystem consists of disjoint spin blocks, the fermion entanglement is not equivalent to the spin entanglement, but the effect is expected to be subleading~\cite{CF:10}.

\section{The stationary state} %

In a uniform setting, at large times after a quantum quench, the stationary 
entanglement entropy of a large subsystem $A=[0,\ell]$ must have the same density as the 
thermodynamic entropy of the statistical ensemble which the system locally 
relaxes to~\cite{Gura13, CoKC14,Fago13,SaPR11, bam-15, dls-13, AlCa17, CCsemiclassics,kaufman-2016}. 
Accordingly,  in the specific case of  noninteracting fermions, the entanglement entropy behaves asymptotically as 
\be
\frac{S_{[0,\ell]}}{\ell}= \frac{S_{YY}[n]}L+o(\ell)\,.
\ee
Here $n(p)=\langle \psi|\hat n_p |\psi\rangle$ is the conserved momentum distribution function and the functional $S_{YY}[n]$ is the Yang-Yang entropy (i.e. the thermodynamic entropy of the statistical ensemble characterised by $n(p)$) 
\be
S_{YY}[n]\equiv L \int {\rm d}p\, s_{\rm YY}\left[n(p)\right]\,,
\label{eq:stat}
\ee
where 
\be
2\pi s_{\rm YY}[n]= -n\ln n -\left(1-n\right)\ln (1-n)\,,
\label{YY}
\ee
is the standard entropy of a fermionic mode which is occupied with probability $n$ and empty with probability $1-n$.

\section{Junction of two stationary states}
The knowledge of the entanglement in a uniform stationary state \eref{eq:stat} is all we need to describe the entire dynamics 
(in the scaling regime) after the junction of two stationary states. 
More generically, the following reasoning applies to all quasistationary initial states, namely those in which the reduced density matrix 
of an arbitrary subsystem commutes with the Hamiltonian, up to boundary terms. 
In the framework of GHD, such states, in the scaling limit, can be  (almost) completely characterised~\cite{Fago17-higher}
by a space and time dependent momentum distribution function $n_{x,t}(p)$ which evolves in time according to the continuity 
equation \cite{CaDY16, BCDF16}
\be
\partial_t n_{x,t}(p) + v(p) \partial_x n_{x,t}(p)=0\,.
\label{eq:continuity_eq}
\ee
This setup is very close to  the classical concept of local equilibrium, where the entropy is additive. It is then natural to assume additivity. We can obtain the entanglement entropy of a finite region by slicing it and summing the contribution \eref{eq:stat} of every slice; this gives   
\be
S_{[x_1,x_2]}(t)= \int_{x_1}^{x_2}\!\!\!{\rm d} x\int {\rm d}p\, s_{\rm YY}\left[n_{x,t}(p)\right]\,.
\label{eq:stat2}
\ee 
The additivity hypothesis can be proven using the exact formalism of Ref.~\cite{Fago17-higher}. In particular, one can check that additivity holds at the first order in the inhomogeneity. We leave a rigorous derivation of  \eref{eq:stat2} to future (more technical) investigations. Here we rely on the physical assumption of additivity and check the prediction against exact numerical data. Note that the formalism of Ref.~\cite{Fago17-higher} allows for a controlled expansion in the inhomogeneity of the state, and can also be used to compute the leading corrections.

\begin{figure}
\begin{center}
	\includegraphics[width=0.6\textwidth]{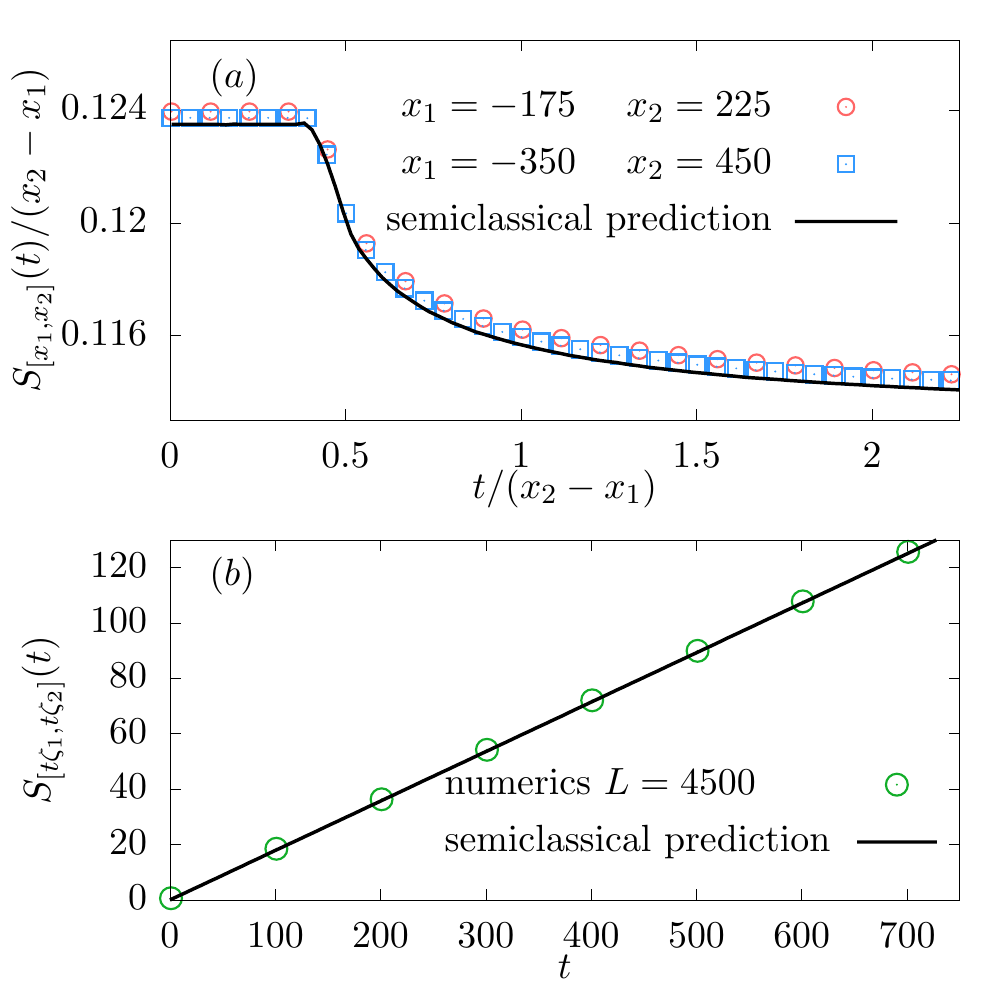}
	\caption{Entanglement entropy  after an inhomogeneous quench in the transverse field Ising chain \eref{eq:H} with $h=3$.  
	The initial state is obtained by joining together two thermal states at different temperatures $T_{\rm L}=1$ and $T_{\rm R}=2$.  
	Points are the result of exact numerics for a system of size $L=4500$, while the lines represent 
	our analytical prediction \eref{eq:stat2}. 
(a): $S_{[x_1, x_2]}(t)/(x_2-x_1)$ versus $t/(x_2-x_1)$	
The numerical data approach the prediction in the limit $x_2\rightarrow\infty$ for fixed $x_1/x_2$.
(b): $S_{[t\zeta_1, t\zeta_2]}(t)$ as function of $t$.
Here we consider $\zeta_1=-0.5$ and $\zeta_2=1$  ($v_{\rm max}=2$ with our conventions).
Note the presence of a small constant offset in the data due to $O(t^0)$ terms. }
	\label{fig1}
	\end{center}
\end{figure}

We tested the validity of Eq.  \eref{eq:stat2} against exact numerical computations in the Ising chain prepared in the 
initial state obtained by joining together two thermal states at different temperatures.
Our prediction is in excellent agreement with numerical simulations, as shown in Fig.~\ref{fig1}. 
The entropy remains constant until the fastest quasiparticles coming from the junction reach one boundary of the subsystem (cf. Fig.~\ref{fig1} (a)); this is a simple example of light-cone effect in the entanglement propagation~\cite{CCsemiclassics}.

\section{Bipartite entanglement in homogeneous settings}  
We summarise here the quasiparticle approach of Ref. \cite{CCsemiclassics} for the entanglement entropy dynamics 
of a one-dimensional model initialised in a homogeneous low-entangled (i.e. subextensive) non-equilibrium state. 
This is not an ab initio calculation but rather a simplified view that captures the physical pith. 
Within this picture the initial state is interpreted as a collection of  {\it pairs} of semiclassical quasiparticle excitations, 
namely entangled particles with definite momentum and position and which follow classical linear trajectories. 
Because of the subextensive initial entanglement, one assumes that only pairs of particles emitted from the same point at ${t=0}$ are
entangled. The quasiparticles move ballistically, and, in the scaling limit $t\sim \ell$, the entanglement entropy is obtained by
summing the contributions of the pairs with one quasiparticle in $A$ and the other in the complement. 
The resulting entropy has the following form \cite{CCsemiclassics}
\bea
S_{[0,\ell]}(t) &=& \int {\rm d}p\, {\rm min}(\ell, 2 |v(p)|t) f(p)
\,,\qquad 
\label{eq:ground}
\eea
where $v(p)=\varepsilon'(p)$ is the group velocity of the quasiparticles with momentum $p$. We point out that this expression  is not yet predictive, as it depends on the unknown function $f(p)$, associated with the entropy density carried by the pair $(p,-p)$. 
In few cases, it was determined from ab-initio calculations, as \emph{e.g.} done for the Ising chain in~\cite{FaCa08}.

It has been pointed out in \cite{AlCa17} that the function $f(p)$ may be read out 
from the property that the entanglement entropy \eref{eq:ground} for $t\to\infty$ reduces to \eref{eq:stat},  so that 
\be
f(p)=s_{\rm YY}\left[n(p)\right].
\label{eq:unknownfunction}
\ee
The crucial assumptions behind~\eref{eq:ground} (with \eref{eq:unknownfunction}) are that entangled quasiparticles are produced in 
pairs and from the same spatial point. Indeed, within the same assumptions in \cite{AlCa17} a similar conjecture has been proposed and tested for the case of generic 
interacting integrable models. 
For free systems, there are only a few examples where these assumptions have been weakened. 
In \cite{CEF:Isinglong1,leda14}, the case of initial states with extensive entanglement entropy has been considered, while,
in \cite{BeTC18}, the initial states produce $n$-plets of correlated excitations with $n>2$. 
In these generalised cases the semiclassical picture is still applicable, but Eq. \eref{eq:ground} must be modified in order to 
catch the more complicated quasiparticle structure of the initial state. 
As the most important difference, the thermodynamic entropy of the stationary state (which is always of the form \eref{YY})
is not enough to fix the entanglement entropy.

\section{Revisiting the semiclassical picture}
The main idea of Ref. \cite{AlCa17} was to reconstruct the entanglement entropy going back in time from its asymptotic value at $t=\infty$.
This idea hardly adapts to the case of inhomogeneous quenches where more information about the initial state is necessary. We propose here an alternative approach which,  going forward in time, allows us to have the entire time evolution
for any structure of correlated quasiparticles. Moreover, it is easily adapted to the inhomogeneous case.

Let us explain this approach for the well known case of an initial pure state which can be expressed in terms of pairs of excitations 
of opposite momenta. Such initial state may be written  (ignoring all unimportant factors/phases/etc) as
\be
|\Psi_0\rangle= \prod_k \Big(\sqrt{1-n(k)}+ i \sqrt{n(k)} b^\dag_k b^\dag_{-k} \Big)|0\rangle,
\label{inipairs}
\ee
corresponding to the density matrix $\boldsymbol \rho_0=\bigotimes_k\boldsymbol \rho_{k,-k}$, where we introduced 
\be
\fl\boldsymbol \rho_{k,-k}\equiv (1-n(k))b_k b^\dag_k b_{-k}b^\dag_{-k}+n(k) b^\dag_k  b_k b^\dag_{-k} b_{-k}+ i \sqrt{n(k)(1-n(k))} (b^\dag_k  b^\dag_{-k}-b_{-k}b_k).
\label{eq:pairdm}
\ee
Incidentally, in Refs \cite{delfino-14, PiPV17} these have been identified as ``integrable initial states".
Eq. \eref{inipairs} encodes all the quantum information about the state, but now we want to understand it semiclassically, i.e. 
in terms of pairs of quasiparticles with definite positions and momenta. 
Since we are dealing with a homogeneous quench, the pairs of quasiparticles must be uniformly distributed in the system. 
Furthermore, the state is factorised in  momentum space and all the entanglement must be only between particles of 
opposite momenta. The reduced density matrix of the particle of momentum (say) $+k$, after having integrating over the one with $-k$ is 
\be
	\boldsymbol \rho_{k}={{\rm Tr}_{-k}[\boldsymbol \rho_{k,-k}]}=
	1- \hat n_{k}+n(k)(2 {\hat n}_{k}-1)\,,
	\label{eq:final_rho}
\ee 
and hence the entanglement entropy between the two particles is 
$-{\rm Tr}[\boldsymbol \rho_{k} \log \boldsymbol \rho_{k}]=2\pi s_{\rm YY}[n(k)]$.
The crucial assumption of the semiclassical picture is that, for $t>0$, quasiparticles move along straight lines with
no interaction and no entanglement generation in momentum space. 
The  entanglement growth in real space is entirely due 
to the spreading of pairs that, moving ballistically, entangle regions farther and farther apart. 
Thus, we conjecture that the reduced density matrix of one interval of length $\ell$ coincides with the reduced density matrix of the quasiparticles that at a given time are within such interval. 
Such reduced density matrix is obtained by means of the following heuristic argument. First, we subdivide the system in cells of size $\Delta $, sufficiently larger than any microscopical length scale (such as the lattice spacing) so that semiclassical trajectories are well defined, but still smaller than the macroscopical scales $\ell$ and $2v(k)t$.
Then, we can \emph{suggestively} write the semiclassical initial density matrix as 
\be
\boldsymbol \rho^{\rm sc}(0)=  \bigotimes_{i=1}^{L/\Delta }\bigotimes_{\tilde k} \tilde {\boldsymbol \rho}_{\tilde k,-\tilde k}\,,
\label{eq:scinit}
\ee
where $\tilde {\boldsymbol \rho}_{ k,- k}={\boldsymbol \rho}_{ k,- k}$ (\emph{cf}. Eq.~\eref{eq:pairdm}) and the momenta $\tilde k$ are defined in the cell. 
The semiclassical reduced density matrix of $A=[0,\ell]$ at time $t$ is obtained following the motion of the quasiparticles; it is given by
\be
\boldsymbol \rho_{A}^{\rm sc}(t)= \bigotimes_{\tilde k} {\boldsymbol \rho}_{\tilde k}^{\otimes \min(2|v({\tilde k})|t,\ell)/\Delta } 
\otimes \boldsymbol \rho_{\rm pure} \,.
\label{eq:redrhosc}
\ee
Here  $\bigotimes_{\tilde k} {\boldsymbol \rho}_{\tilde k}^{\otimes \min(2|v({\tilde k})|t,\ell)/\Delta }$ comes from all pairs of quasi-particles with one particle inside $A$ and the other outside, while $\boldsymbol \rho_{\rm pure}$ comes from the pairs of quasiparticles within $A$. Since the latter is the density matrix of a pure state, it gives zero contribution to the entropy. To write \eref{eq:redrhosc} we used the form \eref{eq:scinit} of the initial state and the fact that a quasiparticle with momentum $\tilde k$ follows a linear trajectory with velocity $v(\tilde k)$: this implies that at time $t$ only $\min(2|v({\tilde k})|t,\ell)/\Delta$ pairs with momenta $\{\tilde k,-\tilde k\}$ have one and only one quasiparticle in the subsystem.

Using \eref{eq:redrhosc} and taking the thermodynamic limit we find
\be
\fl
	  S_{[0,\ell]}(t)={
	  -{\rm Tr}[\boldsymbol \rho_{A}^{\rm sc}(t)\log \boldsymbol \rho_{A}^{\rm sc}(t) ]}\simeq
	  -\!\!\int \frac{{\mathrm d p}}{2\pi} \min(2 |v(p)| t,\ell) \,\,\,
	{\rm Tr}[\boldsymbol \rho_{p} \log \boldsymbol \rho_{p}]\,,
	\label{eq:genericf_homo}
\ee
where the last equality becomes exact in the scaling limit $L\gg \ell,v(p)t\gg \Delta\gg 1$.
After replacing  $-{\rm Tr}[\boldsymbol \rho_{p} \log \boldsymbol \rho_{p}]$ with its value $2\pi s_{\rm YY}[n(p)]$, this equation reproduces 
\eref{eq:ground} with \eref{eq:unknownfunction}. 

The advantage of this interpretation compared to \cite{AlCa17} is that it is easily adapted to the inhomogeneous case.
Indeed an inhomogeneous pure initial state can be thought of as being of the form \eref{eq:scinit} with a position-dependent $\tilde{\boldsymbol \rho}_{k,-k}$, so the entanglement entropy can be accessed in non-uniform settings simply by  performing
an integral over the spatial profile of the initial density of pairs, as we shall soon see.   
Furthermore, it can also be applied to the case of $n$-plets of entangled quasiparticles considered in \cite{BeTC18}: in essence  one has just to replace ${\boldsymbol \rho}_{ k,- k}$ with the more complicated density matrix of the correlated $n$-plet.

\section{Junction of two non-stationary states}
\begin{figure}[t]
\begin{center}
\includegraphics[width=1.1\textwidth]{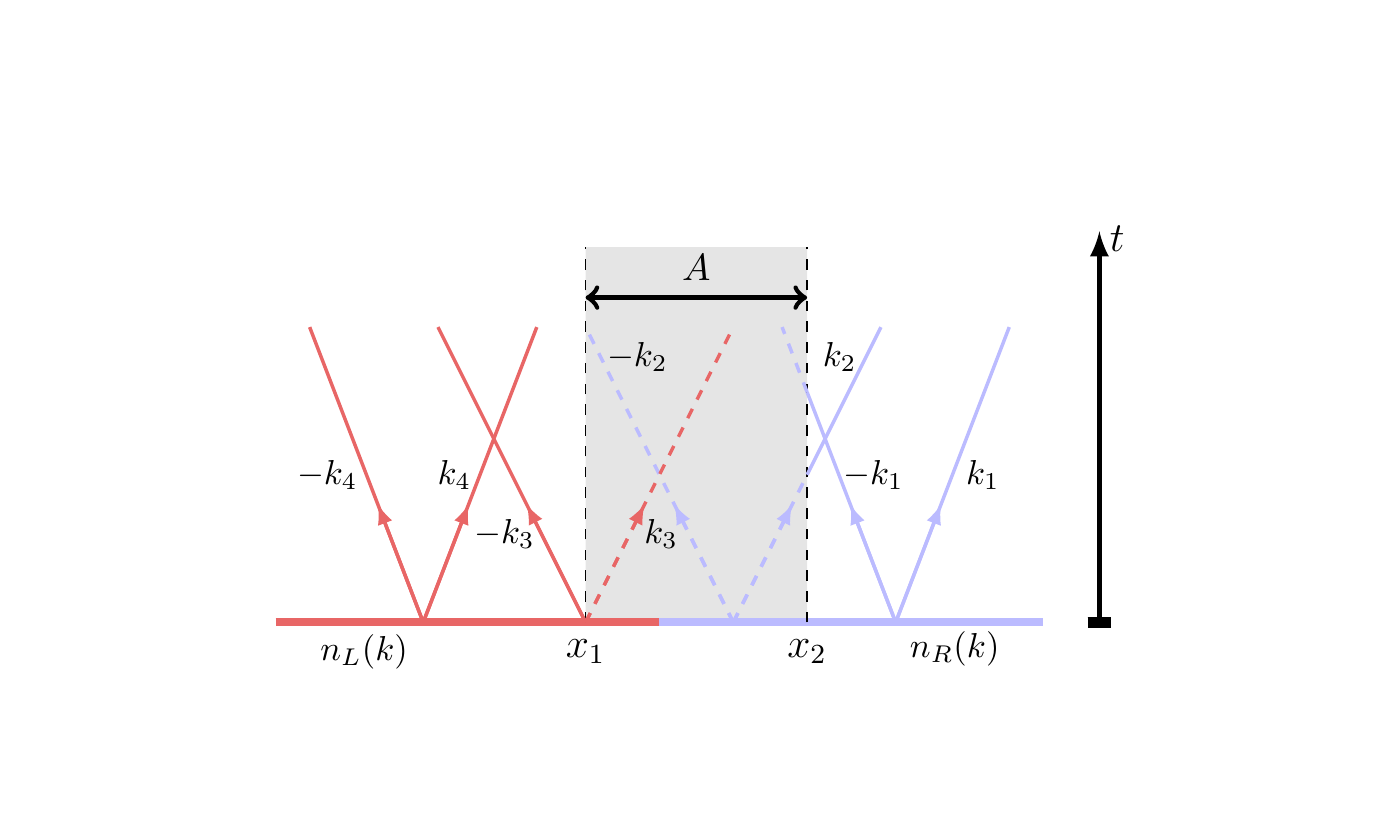}
\caption{Pictorial representation of the motion of the entangled pairs of quasiparticles after the sudden junction of two non-stationary states. The quasiparticles within the subsystem are represented by dashed lines while those out of the system by full lines. Only the pairs with a single particle in the subsystem contribute to the entanglement.}
\label{fig:quasiparticles}
\end{center}
\end{figure}

We now move our attention to the case of an initial state being the junction of two different low-entangled states with the
pair structure \eref{inipairs}.
As long as the initial entropy is subextensive, the picture above can straightforwardly be applied. We can again write the initial state as in \eref{eq:scinit}, the ``local" density matrix $\tilde{\boldsymbol \rho}_{ k,- k}$ is again of the form \eref{eq:pairdm} but the momentum distribution appearing will be $n_R(k)$ if the cell is on the right of the junction and $n_L(k)$ if the cell is on the left of the junction. In other words, the contribution to the entanglement of each quasiparticle pair now depends on whether the pair is originated on the left or on the right of the junction, see Fig.~\ref{fig:quasiparticles}. 

Computing the reduced density matrix of the particles in the system as above, we find the following expression for the entanglement entropy in the thermodynamic limit
\bea
\fl S_{[x_1,x_2]}(t)=\!\int\!
{\mathrm d p}\, \Theta(-v(p))\int_{\max(x_2+2 v(p)t,x_1)]}^{x_2} \hspace{-2cm}{\rm d} x\, f_{x-v(p)t}(p)+\!\int\!
{\mathrm d p}\, \Theta(v(p))\int^{\min(x_1+2 v(p)t,x_2)}_{x_1} \hspace{-2cm}{\rm d} x \, f_{x-v(p)t}(p)\,,
\label{eq:genericf}
\eea
where we introduced a position-dependent entropy density in momentum space
\be
f_x(k)=\Theta(x)s_{\rm YY}[n_{R}(k)]+\Theta(-x)s_{\rm YY}[n_{L}(k)]\, .
\ee
\begin{figure}[t]
\begin{center}
\includegraphics[width=0.6\textwidth]{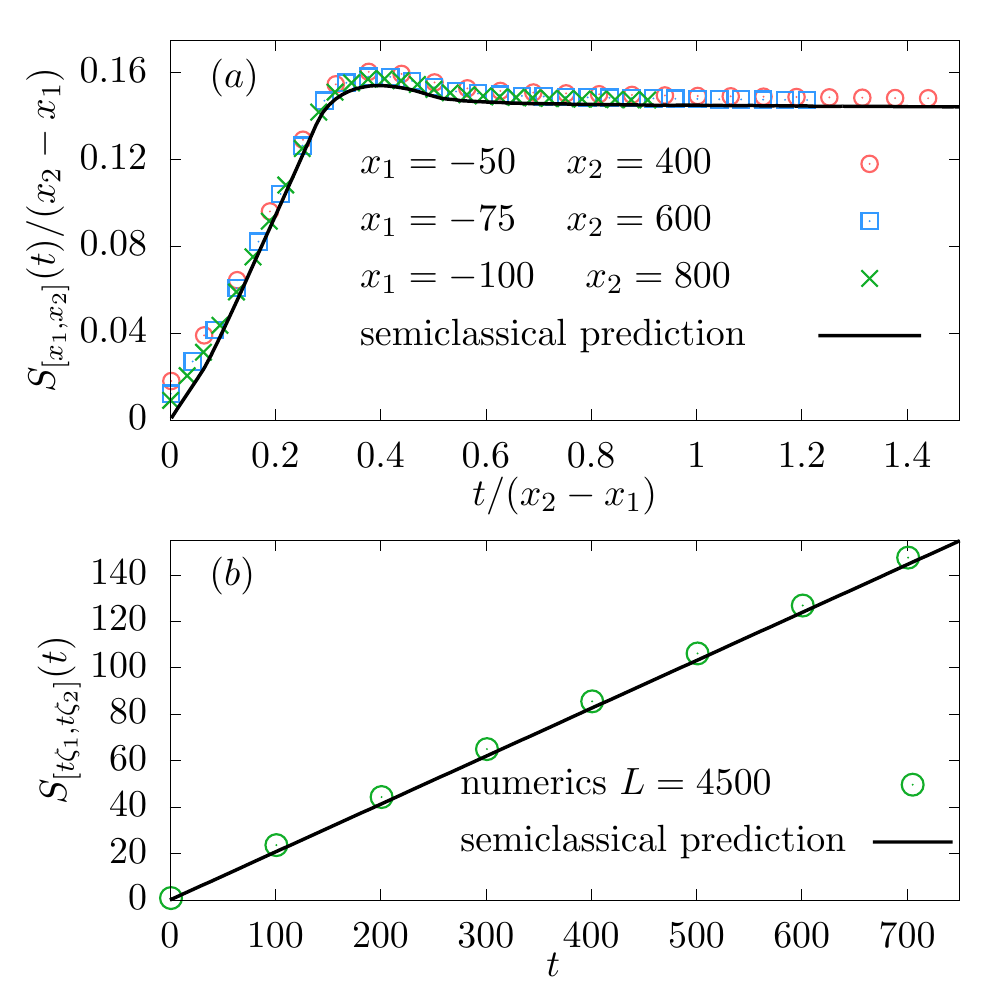}
\caption{Entanglement entropy  after an inhomogeneous quench in the transverse field Ising chain. 
The initial state is obtained by joining together two ground states of the TFIC Hamiltonian \eref{eq:H} for different values of the magnetic field; specifically on the right we have $h=3.5$, on the left $h=0.9$, and the time evolution is performed with the same Hamiltonian for $h=3$. Points are the result of exact numerics for a system of size $L=4500$, while the lines are obtained using our analytical prediction \eref{eq:resultray}. 
(a): $S_{[x_1, x_2]}(t)/(x_2-x_1)$ versus $t/(x_2-x_1)$.
The numerical data approach the prediction in the limit $x_2\rightarrow\infty$ at fixed $x_1/x_2$.(b): $S_{[t \zeta_1, t \zeta_2]}(t)$ versus $t$ with $\zeta_1=-0.5$ and $\zeta_2=1$ ($v_{\rm max}=2$ with our conventions). 
There is a subleading constant offset in the data due to $O(t^0)$ terms.}
\label{fig2}
\end{center}
\end{figure}
The two terms on the r.h.s. of \eref{eq:genericf} are the contributions to the entanglement entropy given by pairs with one quasiparticle in the subsystem and the other outside, on its right and on its left respectively. 

Since the Yang-Yang entropy satisfies a continuity equation of the form \eref{eq:continuity_eq}, we find
\bea
\fl f_{x-v(k) t} &=& \Theta(x-v(k) t) s_{YY}[n_R(k)] +  \Theta(v(k) t-x) s_{YY}[n_L(k)] \nonumber \\ \fl
 &=& s_{YY}[\Theta(x-v(k) t) n_R(k) +  \Theta(v(k) t-x) n_L(k)]= s_{YY}[n_{x,t}(k)],
\eea
and, in turn,
\bea
\fl S_{[x_1, x_2]}(t)=&& \int
{\mathrm d p}\, \Theta(-v(p)) \hspace{-0cm} \int^{{x_2}}_{\max({x_2}+2 v(p)t,{x_1})} \hspace{-2.5cm} \mathrm d x \,s_{\rm YY}\left[n_{x,t}(p)\right]+\int
{\mathrm d p}\, \Theta(v(p))\hspace{-0cm} \int_{{{x_1}}}^{\min({x_1}+2 v(p)t,{x_2})} \hspace{-2.5cm}\mathrm d x \,s_{\rm YY}\left[n_{x,t}(p)\right]\,.
\label{eq:resultray}
\eea
As a check, we note that this expression recovers the results obtained in the conformal case~\cite{WWR18}. Moreover, we tested it against exact numerics for systems up to $L=4500$ finding a good agreement, \emph{cf}. Fig.~\ref{fig2}. 
The example shown in Fig.~\ref{fig2} (a) is worth a comment. We see that the entanglement entropy is not a monotonous function of time. This situation can not be realised in the homogeneous case, as one can readily infer from \eref{eq:ground} and \eref{eq:unknownfunction}. On the contrary, this can happen after the junction of two different states when the subsystem is deep on the side with larger Yang-Yang entropy. In this case the initial growth of $S_{[x_1, x_2]}(t)$ is due to pairs coming from the high-entropy side, while the arrival of quasiparticles coming from the other side lowers the entanglement.

\section{Conclusions}
We have studied the evolution of the entanglement entropy after inhomogeneous quenches in generic systems of free fermions. 
We focused on the protocol where two semi-infinite systems, initially prepared in different macrostates, are suddenly joined together and left to evolve unitarily. We computed the entanglement between an arbitrary interval and the rest of the system, obtaining analytic formulae in two prototypical examples: (i) the two initial macrostates are stationary; (ii) the two initial macrostates are non-stationary pure states producing only pairs of quasiparticles.
Our formulae display a number of interesting features: most prominently, we find that the entanglement dynamics is not uniquely specified by the thermodynamic information given by the generalised hydrodynamic approach. To make quantitative predictions one needs a larger amount of microscopic information about the structure of the initial state.

Although the semi-classical picture employed to derive our results is non-rigorous, we expect our formulae to become exact in the limit of infinite time: on the one hand, we tested our findings to extremely high numerical precision; on the other hand, we believe that a rigorous proof of the latter should be possible (even though non-trivial) by employing a combination of the techniques used in \cite{FaCa08} and \cite{Fago17-higher}. 

All our results can directly be adapted to bosonic noninteracting models by replacing the Yang-Yang
entropy with the appropriate bosonic analogue, as in the homogeneous case \cite{AlCa17,p-18}. 
Furthermore, for free models the same arguments lead to exact formulae also for the R\'enyi entanglement entropy of arbitrary order, 
but this ceases to be the case for interacting systems \cite{ac-17b}.

Having established the validity of the quasiparticle picture for the computation of entanglement from inhomogeneous settings in the free case, it is natural to wonder whether generalisation to the interacting integrable case is possible; this issue is currently under investigation~\cite{ABF:18}.

\ack{ We thank Vincenzo Alba for useful discussions. B. B. acknowledges financial support by the ERC under the Advanced Grant 694544 OMNES. B. B. and M. F. thank SISSA for hospitality. Part of this work has been carried out during the workshop ``Quantum paths'' at the Erwin Schr\"odinger International Institute for Mathematics and Physics (ESI) in Vienna.}

\appendix

\section{Numerical Method}
\label{app:num}

The time evolution of the entanglement entropy of one block of contiguous sites in free fermionic systems 
(or systems mappable to free fermions)
evolving from gaussian states can be computed exactly using an efficient numerical procedure which we now describe. The Hamiltonian of such systems can be (effectively) written in the following quadratic form
\be
H=\frac{J}{4}\sum_{n,l=1}^L
\left(\begin{array}{cc}
a_{n-\frac{L}{2}}^x & a_{n-\frac{L}{2}}^y
\end{array}\right)
\mathcal H_{n,l}
\left(\begin{array}{c}
a_{l-\frac{L}{2}}^x \\
a_{l-\frac{L}{2}}^y
\end{array}\right)\,,
\label{eq:HMajorana}
\ee
where $J$ sets the relevant energy scale, $\{\mathcal H_{n,l}\}$ are ${2\times2}$ matrices, and the Majorana fermions $a^x_n, a^y_l$ satisfy $\{a^\alpha_n,a^\beta_l\}=2\delta_{nl}\delta_{\alpha\beta}$. For definiteness we considered a system defined on a lattice of $L$ sites.

As Wick's theorem applies at all times, the entanglement entropy of the subsystem $A= \{x_1,\dots, x_2\}$ with respect to the rest of the chain is completely specified by the two-point correlations of Majorana fermions as follows~\cite{CCsemiclassics, FaCa08, pe-09, lrv-03}
\be
S_A(t)= -{\rm Tr}\left[\left(\frac{I_{2|A|}-\Gamma_A(t)}{2}\right) \ln\left(\frac{I_{2|A|}-\Gamma_A(t)}{2}\right)\right]\,.
\label{eq:entropy}
\ee
Here $I_n$ is the $n$-by-$n$ identity, and $\Gamma_A(t)=P_{A} \Gamma(t) P_A^t$, where $P_A$ is the $2|A|$-by-$2L$  matrix with $[P_A]_{\ell, n}=\delta_{\ell+2 x_1,n}$  and the correlation matrix $\Gamma(t)$ reads as  
\be
\Gamma(t)=\left(
\begin{array}{ccc}
[\Gamma_{2}(t)]_{1,1} & \ldots & [\Gamma_{2}(t)]_{1, {L}}\\
\vdots & \ddots & \vdots\\
\,[\Gamma_{2}(t)]_{L, 1} & \ldots & [\Gamma_{2}(t)]_{L, L}
\end{array}\right)\,,
\label{eq:Gammamat}
\ee
with 
\be
\!\!{\left[\Gamma_{2}(t)\right]}_{n,m}=\delta_{n,m} I_2 -\left(\begin{array}{cc}
 \mathrm{Tr}[\boldsymbol\rho_t a_{n-\frac{L}{2}}^x a_{m-\frac{L}{2}}^x ] &\mathrm{Tr}[\boldsymbol\rho_t a_{n-\frac{L}{2}}^x a_{m-\frac{L}{2}}^y ] \\
 \mathrm{Tr}[\boldsymbol\rho_t a_{n-\frac{L}{2}}^y a_{m-\frac{L}{2}}^x ] &\mathrm{Tr}[\boldsymbol\rho_t a_{n-\frac{L}{2}}^y a_{m-\frac{L}{2}}^y ] 
\end{array}\right)\,.
\label{eq:appCorrmat}
\ee
Here $\boldsymbol\rho_t$ is the state of the system at time $t$. 

Using the algebra of the Majorana fermions, it is evident that the $2L\times2L$ matrix $\mathcal H$, formed placing the blocks $\mathcal H_{n,l}$ as in \eref{eq:Gammamat}, can be chosen to be skewsymmetric, so that the correlation matrix $\Gamma(t)$ can be written in terms of $\Gamma(0)$ as follows (see Ref.~\cite{F:current} for further details)
\be
\Gamma(t)=e^{-i \mathcal H t}\Gamma(0)e^{i \mathcal H t}\,.
\label{eq:corrmat}
\ee
The time evolution of the entanglement entropy is determined in three steps. (i) Define the initial correlation matrix $\Gamma(0)$. To do that, it is convenient to introduce a smoothened version of the initial state that interpolates between the states on the two sides of the junction over a finite length-scale $\lambda=O(t^0)$; we use such state in Eq.~\eref{eq:appCorrmat} for $t=0$ to produce the initial correlation matrix. (ii)  Evolve $\Gamma(0)$ using \eref{eq:corrmat} for a set of discrete times $\{t_i\}_{i=1,\ldots,N}$. (iii) For each $t_i$, diagonalise $\Gamma_A(t_i)$ numerically and evaluate \eref{eq:entropy}. This procedure works efficiently for very large systems $L\sim 5000$ up to very large times $J t_{\rm max}\sim 800$ and can be implemented to arbitrary precision. 

\section*{References} %

\end{document}